\documentclass[amsmath,amssymb,showpacs]{revtex4}

\usepackage{graphicx}
\usepackage{dcolumn}
\usepackage{bm}
\usepackage{epsfig}
\usepackage{color}
\usepackage{colordvi}
\usepackage[dvipsnames]{xcolor}
\usepackage{relsize}
\usepackage{accents}
\usepackage{wasysym}  

\newcommand{\be}{\begin{equation}}
\newcommand{\ee}{\end{equation}}
\newcommand{\ba}{\begin{eqnarray}}
\newcommand{\ea}{\end{eqnarray}}
\newcommand{\n}{\nonumber}
\newcommand{\lan}{\langle}
\newcommand{\ran}{\rangle}

\newcommand{\p}{\partial}

\newcommand{\defeq}{\mathrel{\mathop:}=}

\begin{document}
\title{On phase space representation of non-Hermitian system with $\mathcal{PT}$-symmetry}

\author{Ludmila Praxmeyer$^1$, Popo Yang$^2$, and Ray-Kuang Lee$^{1,2}$}

\affiliation{
$^1$ Institute of Photonics Technologies, National Tsing-Hua University, Hsinchu, Taiwan\\
$^2$ Department of Physics, National Tsing-Hua University, Hsinchu, Taiwan}

\begin{abstract}
We present a  phase space study of non-Hermitian Hamiltonian with
$\mathcal{PT}$-symmetry based on the Wigner distribution function.
For an arbitrary complex potential, we derive a generalized
continuity equation for the
 Wigner function flow and calculate the related circulation values.
Studying vicinity of an exceptional point, we  show that  a
$\mathcal{PT}$-symmetric phase transition from an unbroken
$\mathcal{PT}$-symmetry phase to a broken one  is a second-order
phase transition.
\end{abstract}


\maketitle

\section{Introduction}
With spatial reflection and time reversal,  parity-time
($\mathcal{PT}$)-symmetry has a special place in studies of
non-Hermitian operators, as it reveals the possibility to remove
the restriction of Hermiticity from Hamiltonians.
A $\mathcal{PT}$-symmetry Hamiltonian can exhibit entirely real and positive eigenvalue spectra~\cite{bender, bender2, bender3}. Even though the attempt to
construct a complex extension of quantum mechanics was ruled out
for violating the no-signalling principle when applying the local
$\mathcal{PT}$-symmetric operation on one of the entangled
particles~\cite{YiChan}, such a class of non-Hermitian systems are
useful as an interesting model for open systems in the classical
limit. Through the equivalence between quantum mechanical
Schr\"odinger equation and optical wave equation, with the
introduction of a complex potential, $\mathcal{PT}$-symmetric
optical systems demonstrate many unique features. In
$\mathcal{PT}$-symmetric optics, wave dynamics is not only
modified in the linear systems, such as synthetic optical
lattices~\cite{makris2008, Guo} and waveguide
couplers~\cite{Ruter, coupler}, but also in the nonlinear
systems~\cite{soliton}.

A Hamiltonian $\hat{H}$ is $\mathcal{PT}$-symmetric if it commutes
with the $\mathcal{\hat{P}\hat{T}}$ operator,
{\mbox{$[\mathcal{\hat{P}\hat{T}}, \hat{H}] = 0$}}. Here,
$\mathcal{\hat{P}}$ is the spatial reflection operator that takes
$x\rightarrow -x$; while $\mathcal{\hat{T}}$ is the time reversal
anti-linear operator that takes $i\rightarrow -i$. One can easily
check that the eigenvalues of $\hat{H}$ are always real when the
eigenstates of a $\mathcal{PT}$-symmetric Hamiltonian are also the
eigenstates of $\mathcal{\hat{P}\hat{T}}$. It is also known that
there exist spontaneous $\mathcal{PT}$ symmetry-breaking points,
where the eigenstates of $\hat{H}$ are no longer the eigenstates
of $\mathcal{\hat{P}\hat{T}}$. Depending on the $\mathcal{PT}$
symmetry-breaking condition, eigenvalues of a $\mathcal{PT}$
symmetric operator are either real or complex conjugate
pairs. The former scenario is called the unbroken
$\mathcal{PT}$-symmetry phase; while the latter one is known as
the broken phase. The transition point from a unbroken to a broken
$\mathcal{PT}$-symmetry  phase is coined the exceptional point
(EP). By steering the system in the vicinity of an EP,
loss-induced suppression of lasing~\cite{loss} and stable
single-mode operation with the selective whispering-gallery
mode~\cite{lasing} are implemented with state-of-the-art
fabrication technologies.

A natural questions arises about the existence of these
exceptional points and the relevant order of phase transitions. In
this work, we present a phase space study showing how the
symmetry-breaking manifests in systems governed by non-Hermitian
$\mathcal{PT}$-symmetric Hamiltonians using an example of
generalized quantum harmonic oscillators. Even though, from a
physicist point of view, only operators with purely real
eigenvalues are the observables -- an eigenvalue with a non-zero
imaginary part cannot be interpreted as a result of measurement.
For a $\mathcal{PT}$-symmetric Hamiltonian, non-real eigenvalues
always appear in complex conjugate pairs ensuring conservation of
energy in the system.  It distinguishes $\mathcal{PT}$-symmetry
operators from other non-Hermitian operators, making the class
especially interesting. With the introduction of Wigner function
flow, we derive the corresponding continuity equation. Moreover,
through the Gauss-Ostrogradsky theorem, we show that the phase
transition in the vicinity of EP, i.e., from a unbroken
$\mathcal{PT}$-symmetry phase to a broken one, is a continuous
function of the system parameter, which indicates that a
$\mathcal{PT}$-symmetric phase transition is a second-order phase
transition.

\section{Model Hamiltonian for $\mathcal{PT}$-symmetry systems}

From a quantum mechanical perspective,  operators  with complex
eigenvalues cannot be related to observables,  thus, a notion of
{\it non-Hermitian Hamiltonian}  has no place in  orthodox quantum
theory. Nevertheless, an idea of releasing the Hermiticity
requirement for Hamiltonians appears repeatedly in literature
~\cite{bender,bender2,bender3,znojil,znojil2},
 under justification that there exist non-Hermitian operators with purely real spectra.
However, proofs of spectra reality are definitely nontrivial,
 usually even finding domain on which an operator acts can lead to serious difficulties,
 which is often ignored when considering the non-Hermitian Hamiltonians  \cite{comment1}.
In the following,  in order to maintain some level of formal
 rigour and mathematical correctness, we shall talk about finding solutions of
differential equations rather then extending quantum mechanics to
non-Hermitian systems.

Here, we consider a family of differential equations parameterized by a
 continuous parameter $\,\epsilon>0\;$ of the form:
\begin{equation}
\frac{\partial^2 \psi(x)}{\partial
x^2}+V_\epsilon(x)\psi(x)+2E \psi(x)=0,
\label{diff}
\end{equation}
where $E$ is the corresponding eigen-energy, $x$ denotes a real
variable, and $\psi(x)$ is a square integrable function. This
stationary Schr\"odinger wave equation is introduced as a
$\mathcal{PT}$-symmetry system from a generalized quantum harmonic
oscillator~\cite{bender}. Unlike in the case of traditional
quantum mechanics, we allow the potential function, i.e.,
$V_\epsilon(x)$ shown in Eq. (1),  to take on complex values.
 Here, let us specify the definition of $V_\epsilon(x)\defeq -(ix)^\epsilon$,
  by  stating explicitly which branch of logarithm  will be used in this paper:
\ba
V_\epsilon(x)=-(ix)^{\textstyle{\epsilon}}= e^{\textstyle\epsilon
\log(ix)}=
\begin{cases}
       -|x|^{\textstyle{\epsilon}}\big[\cos( {\textstyle{\epsilon}}\frac{\pi}{2})
        +i \sin( {\textstyle{\epsilon}}\frac{\pi}{2} )\big],  & \mathrm{for}\; x>0;\\
           \quad 0, & \mathrm{for}\; x=0;\\
            -|x|^{\textstyle{\epsilon}}\big[\cos( {\textstyle{\epsilon}}\frac{\pi}{2} )
               -i \sin( {\textstyle{\epsilon}}\frac{\pi}{2} )\big],  & \mathrm{for}\; x<0\,.
 \end{cases} \label{potencjal}
\ea
It is easy to notice, that for $\epsilon=2$ this potential
reduces Eq.~(\ref{diff}) to the Schr\"odinger equation of a
quantum harmonic oscillator expressed in the units $m=1$,
$\hbar=1$, $\omega=1$.
In this special case, solutions are given
by Fock (number) states, denoted in the Dirac notation by kets $|n\ran$,
$|m\ran$, etc.
The corresponding eigenfunction of the $n$-th excited state of a quantum harmonic
 oscillator in the  position representation reads
\begin{equation}
u_n(x)=\lan x|n\ran= \frac{1}{\sqrt{2^n n! \sqrt{\pi}}}\, H_n(x)
\, e^{-\frac{x^2}{2}},
\label{hermity}
\end{equation} 
where $H_n(x)$ is the $n$-th order Hermite polynomial. The
corresponding eigenvalues are equal to $E_n=n+1/2\,$, for any
$n\in{\mathbb{N}}$. We are interested in finding pairs
$(\psi_n,\,E_n)_\epsilon\,$ fulfilling Eq.~(\ref{diff}) for  a set
$\epsilon>0$.

\begin{figure}
\begin{center}
 \includegraphics[width=10cm]{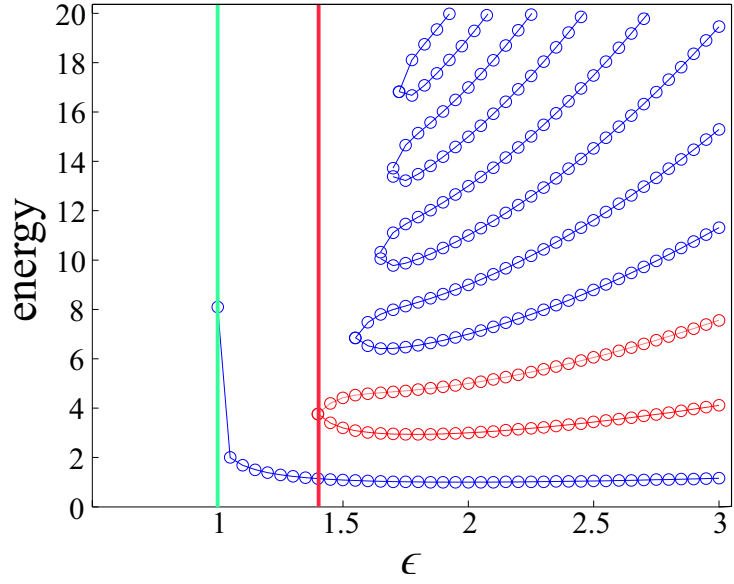}
\end{center}
\caption{Real eigenvalues from the energy spectrum for the
generalized quantum harmonic oscillator from Eqs. (1-2),
  generated as a function of the parameter $\epsilon$.
The vertical line in Red color indicates the exceptional point (EP) at $\epsilon \approx 1.42207$,
where two branches from the $1$st- and $2$nd-excited states marked in Red colors merge together.}
\end{figure}

To find the solutions of the eigenvalue problem with the complex
potential, we use connection with a quantum harmonic oscillator
and solve Eq.~(\ref{diff}) in the Fock state basis. It turns out,
that an analytical formula for the matrix element
$a_{nm}(\!\epsilon)=\lan m|\mathrm{H}_\epsilon|n\ran$ of
$\mathrm{H}_\epsilon=\frac{1}{2}\frac{\partial^2 }{\partial
x^2}+\frac{V_\epsilon(x)}{2}$ can be constructed for any natural
number $n$, $m$ and positive $\textstyle\epsilon$. It reads: \ba
a_{nm}(\!\epsilon) &=& {\textstyle{\frac{\sqrt{n(n-1)}}{4}
\,\delta_{m,n-2} +\frac{\sqrt{(n+1)(n+2)}}{4}\,
\delta_{m,n+2}-\frac{2n+1}{4}\,\delta_{m,n}}}+
 \n\\
 &+&
\Big[{\textstyle{\frac{1-(-1)^{ \widetilde{n} + \widetilde{m}  }  }{4}}}
 \cos( \textstyle{ {\textstyle{\epsilon}} \frac{\pi}{2} })
 + \frac{1+(-1)^{\widetilde{n} + \widetilde{m}   }  }{4}
 i\sin( {\textstyle{\epsilon}}\frac{\pi}{2}) \Big]
 \frac{(-1)^{\lfloor\!\frac{n}{2}\!\rfloor+\lfloor\!\frac{m}{2}\!\rfloor}
  {\textstyle{ 2^{\widetilde{n} + \widetilde{m} }  n! m! }}}{\lfloor\!\frac{n}{2}\!
  \rfloor!\, \lfloor\!\frac{m}{2}\!\rfloor!}\times\n\\
 &&\times
           \;
\Gamma\!\left( {\textstyle{\frac{1+{\textstyle{\epsilon}}+\widetilde{n} +
\widetilde{m}  }{2}}} \right)
 F_{\!\!{}_A}\!\!\left( {\textstyle{\frac{1+
{\textstyle{\epsilon}}+
       \widetilde{n} + \widetilde{m}  }{2};
-\lfloor\!{\textstyle\frac{n}{2}}\!\rfloor ,
-\lfloor\!{\textstyle\frac{m}{2}}\!\rfloor ; {\textstyle
\frac{2\widetilde{n}+1 }{2}}, {\textstyle\frac{2\widetilde{m}+1
}{2}}; 1,1 }}\right)\delta_{m,n}\label{all}\;, \ea where $\Gamma$
is an Euler gamma function; $ F_{\!\!{}_A}$ is a Lauricella
hypergeometric function; symbol $\lfloor\,\rfloor$ denotes a floor
function: $\lfloor \!k\!\rfloor$ is the largest integer not
greater than $k$; character tilde  $\;\widetilde{}\;$ denotes a
binary parity function: $\widetilde{k}$ is 0 for an even $k$ and 1
for an odd $k$. All mathematical formulas needed for derivation of
Eq. (4) are presented in {\bf Appendix A}.

In the analytical expression shown in Eq. (4), the first line
comes from the well known formula $ 2\frac{\partial^2 }{\partial
x^2} \, |n\ran =\sqrt{n(n-1)} \,|n-2\ran
-(2n+1)|n\ran+\sqrt{(n+1)(n+2)}\, |n+2\ran$;  while the next lines
are a combination of parity coefficients and an
Erd\`elyi formula  for the integral $\int_0^{\infty}
   e^{-\lambda x^2} H_{\mu_1}\!(\beta_1 x) H_{\mu_2}\!(\beta_2 x)\cdots
   H_{\mu_n}\!(\beta_n x) x^{\nu} dx\,,$~\cite{Erdelyi}.
Let us note, that crucial to the derivation are parity properties
of Hermite polynomials, e.g., the fact that
 \ba
 \int_{-\infty}^0
 e^{-x^2} H_n(x) H_m(x) |x|^{\textstyle{\epsilon}}dx
 =(-1)^{ \widetilde{n}+\widetilde{m} } \int_0^{\infty}
   e^{-x^2} H_n(x) H_m(x) |x|^{\textstyle{\epsilon}} dx\label{parity0}.
\ea
Whenever $n$ and $m$ have the same parity, the values of
$a_{nm}(\epsilon)$, Eq.~(\ref{all}), are real. Otherwise,
 values $a_{nm}(\epsilon)$ are purely
imaginary. Unless $\epsilon$ is  an even number, matrix
$M_{nn}(\epsilon)$ constructed from elements $a_{nm}(\epsilon)$ is
symmetric ($M=M^\mathrm{T}$) but non-Hermitian. Numerical
diagonalization of $M_{nn}(\epsilon)$, and hence a necessity to
truncate the Hilbert space to a finite basis, sets some formal
limitation on generality of the results. On the other hand, the
method allows for a direct control of precision: truncating basis
at $n_{max}$ we automatically know that eigenfunction are expanded
into a polynomial of the order $n_{max}$, as \ba
\psi_j(x,t)=\sum_{k=0}^{n_{max}} a_{jk} u_k(x) e^{-\frac{i E_k
t}{\hbar}}. \n \ea
 However, one has to realize that, as long as a finite basis is
used, diagonalization of $M_{nn}(\epsilon)$ always leads to
discrete spectra.  In the general case of a non-truncated basis,
there is no way to determine a priori if the spectrum is discrete
or even if there exists a square integrable solution of
Eq.~(\ref{diff}).

Using the matrix elements derived in Eq. (4),
 we  diagonalize the  matrix $M_{nn}(\epsilon)$ numerically, having truncated the Fock basis
 to the first $31$,  $51$, or $71$ elements. For low energy states, already the smallest
 basis of 31 elements gives more than sufficient accuracy.
In Fig.~1, we show real eigenvalues of the energy spectrum
corresponding to the generalized quantum harmonic oscillator
described in Eq. (1). The parameter $\epsilon$ from Eq. (2) is
used as a variable. One can see that  for $\epsilon<2$  the number
of real eigenvalues decreases. When $\epsilon = 1$,  only the
ground state has a real energy;
 while for parameter {\mbox{$1<\epsilon<2$}},  the lowest eigenvalues
of~(\ref{diff}) are real whereas higher eigenvalues might appear
in complex conjugate pairs.
It was conjured that for $\epsilon\geq2$ eigenvalues of Eq. (\ref{diff}) are always real;
however  there is no analytical proof of this statement and although some
 numerical simulations
 support the conjecture, others are inconclusive~\cite{bender}.
In the following, we focus us on the exceptional point (EP) for
the $1$st- and $2$nd- excited states at $\epsilon_{EP}  \approx
1.42207$, as indicated by the vertical Red-colored line in Fig. 1.
At this EP,  two subsequent (up to that point) real eigenvalues
start to have the same absolute value but complex conjugate
imaginary parts. It is a point at which solutions that break
symmetry of the Hamiltonian suddenly appear. In the next Section,
we will examine the phase space representation of such
eigenfunctions and look for a signature of this exceptional point.

\begin{figure}
 \begin{center}
 \includegraphics[width=14cm]{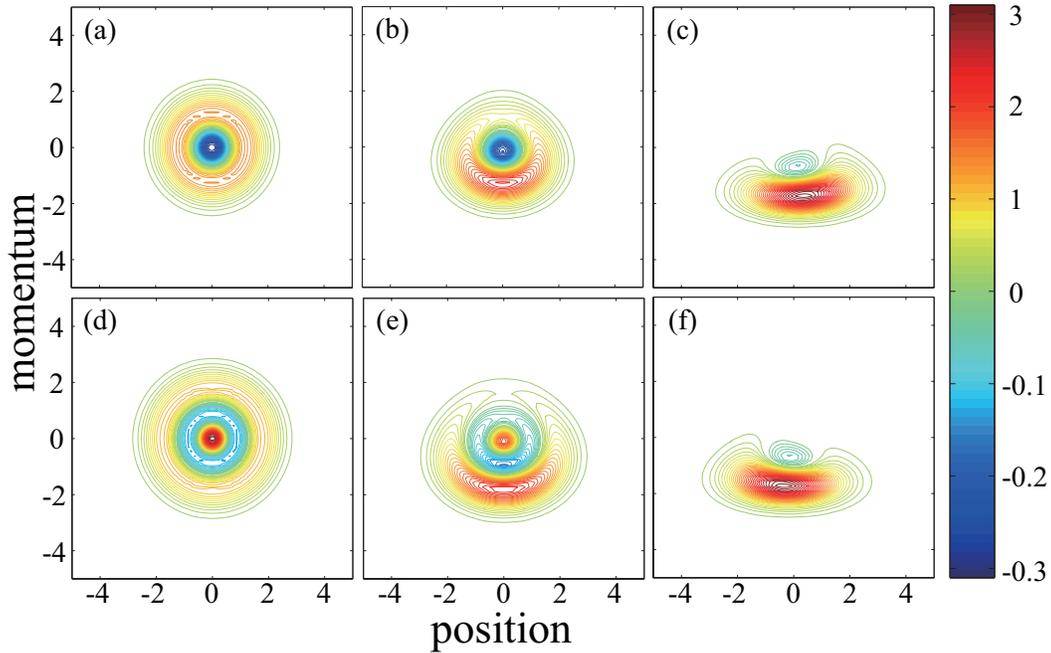}
\end{center}
\caption{Wigner function distributions for the $1$st- (Upper panel) and $2$nd- (Lower Panel)
excited states, at  (a): $\epsilon=2.0$, (b): $\epsilon=1.5$, and
(c): $\epsilon=1.4$. It is noted that the exceptional point appears
at $\epsilon_{EP} \approx 1.42207$.}
\end{figure}

\begin{figure}[t]
 \begin{center}
 \includegraphics[width=14cm]{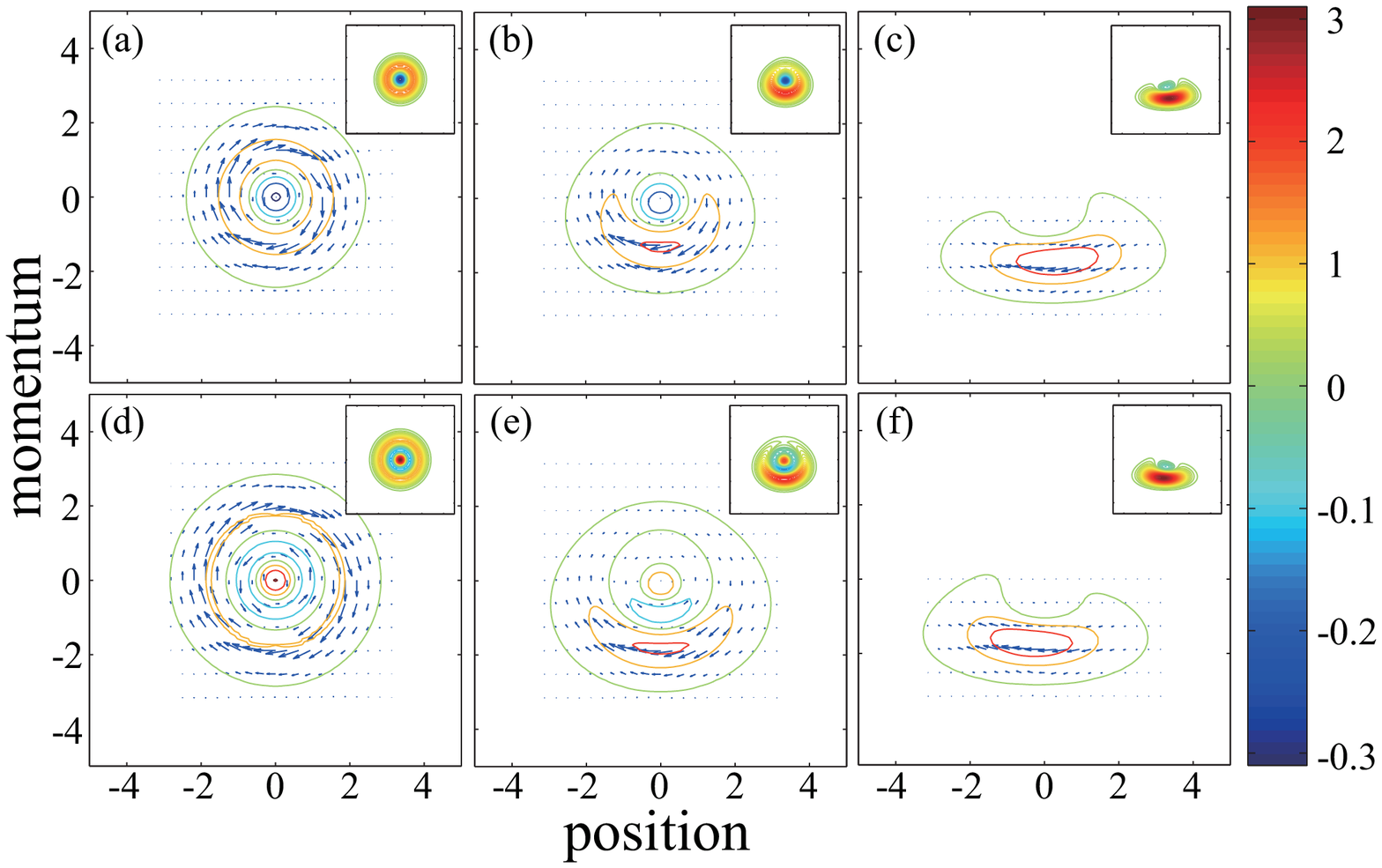}
 \end{center}
\caption{\small  Wigner function currents for the $1$st- (Upper
panel) and $2$nd- (Lower Panel) excited states, at  (a):
$\epsilon=2.0$, (b): $\epsilon=1.5$, and (c): $\epsilon=1.4$.
Here, Arrows denote direction of $\bar{J}_{{\,\psi}}$; while Color
of the arrows depends on the norm $N_{\!{}_{J}}$ with dark gray
denoting the highest value. Again, the exceptional point happens
at $\epsilon_{EP} \approx 1.42207$. Note that depending on the
sign of Wigner function, the rotation may be  clockwise or
counter-clockwise. Insects show the corresponding Wigner function
distributions.}\label{fig4}
\end{figure}

\section{$\mathcal{PT}$-symmetry in Wigner function representation}

Phase space wave characteristics of a square-integrable
wavefunction $\psi(x,t)$ is often examined through the
corresponding Wigner distribution \cite{Wigner, book-phasespace},
defined  as an integral
 \begin{equation}
W_{\!\psi}= W_{\psi}(x,p,t)\defeq
\frac{1}{2\pi \hbar} \int
\psi^*(x+{\textstyle{\xi}}/2,t) \psi(x-{\textstyle{\xi}}/2,t)
\,e^{i {\xi} p/\hbar}\,d{{\xi}},\label{wig}
 \end{equation}
 where $\psi^*$ denotes a complex conjugate of $\psi$.
The Wigner function is always real. It is also normalized to $1$ for
any normalized $\psi$, but in contrast to proper probability
distributions it might take on negative values. The  Wigner
representation
 reflects proper probability properties in position and
momentum representations  simultaneously, and thus is very well
suited to reveal the  possible symmetries in
wavefunctions~\cite{book-phasespace, bauke}.

In Fig. 2, the Wigner  function corresponding to the $1$st- and
$2$nd-exited states of Eq.~(1) with a $\mathcal{PT}$-symmetric
potential given in Eq.~(2) are plotted, for $\epsilon = 2.0$,
$1.5$, and $1.4$, respectively. We start with the case of a
quantum harmonic oscillator ($\epsilon=2$), which has all
eigenvalues real and the corresponding Wigner functions have a
cylindrical symmetry, as shown in Fig. 2(a) and (d). The Wigner
function of a harmonic oscillator depends on $r=\sqrt{x^2+p^2}$ as
a $n$th-order Laguerre polynomial suppressed by exponential
factor. When $\epsilon\neq 2$, such a cylindrical symmetry
vanishes. It is noted from Fig. 1 that the exceptional point for
these two states happens at $\epsilon_{EP} \approx 1.42207$.
Within the unbroken $\mathcal{PT}$-symmetry phase, $\epsilon >
\epsilon_{EP}$, we have the real eigenvalues and the corresponding
Wigner functions  are symmetric under transformation $x
\rightarrow  -x$, as shown in Fig. 2(b) and (e) for the $1$st and
$2$nd exited states, respectively. However, at the same time, the
symmetry $p \rightarrow -p $ that is present in a quantum harmonic
oscillator case disappears. Moreover, as expected, the bigger
difference between the value of $\epsilon$ and $2$, the less
similarity of Wigner distributions to those of a quantum harmonic
oscillator is exhibited.

When the $\mathcal{PT}$-symmetry is broken for $\epsilon <
\epsilon_{EP}$, the corresponding eigenvalues both for the $1$st
and $2$nd exited states not only have non-zero imaginary parts, but
also form a complex conjugate pair to each other. In Fig. 2(c) and
(f), we plot the Wigner function distributions in such a
$\mathcal{PT}$-symmetry-broken phase. As one can see,  neither
symmetry $p \rightarrow -p $ or $x \rightarrow -x$ is valid.
However,  this pair of eigenfunctions with the complex conjugates in
their eigenvalues are mirror images to each other, with respect to
$x = 0$. The same mirror-image symmetry holds for any pair of
eigenfunctions that have complex conjugate eigenvalues.

In addition to the Wigner function distribution in the phase
space, we also introduce a Wigner function flow $\bar{J}_{\psi}$
defining the field $\bar{J}_{\psi}=(J_x,J_p)$ as
\begin{eqnarray}
J_x &=&\frac{p}{m}\, W_{\!\psi}, \,\label{jx}\\
J_p &=& -\sum_{j=1}^\infty
\frac{ (-i\hbar)^{j-1} }{ j!\, 2^j} \bigg[
\frac{d^j V^{\!*}}{dx^j}+(-1)^{j-1} \frac{d^j
V}{dx^j}\bigg] \frac{\p^{j-1} W_{\!\psi}}{\p p^{j-1}},
\end{eqnarray}
where $V$ denotes a potential from the Schr{\"o}dinger equation
that led to eigenfunction $\psi$,  and $V^*$ is its complex
conjugate. In general, a continuity equation
is given by
\ba
&& \frac{\partial W_{\!\psi}}{\partial
  t} + \frac{\p J_x}{\p x}
 +\frac{\p J_p}{\p p}
=\frac{i}{\hbar}\big(V^{\!*}-V
\big)
W_{\!\psi}.
\label{cont}
\ea
In the Hermitian case, the right-hand-side of Eq.~(\ref{cont})
vanishes and in the classical limit of $\hbar\rightarrow 0\;$
it is reduced to
\ba
&& \frac{\partial W_{\!\psi}}{\partial t} +\frac{p}{m}\,\frac{\p
W_{\!\psi}}{\p x}
 -\frac{\p V }{\p x}\,\frac{\p W_{\!\psi}}{\p p}=0.
\label{contClassic}
  \ea
This well-known formula determines the classical evolution of the
Wigner function. It was first derived by Wigner~\cite{Wigner} and
later discussed, e.g., by Wyatt~\cite{Wyatt}.
An analogous equation  valid when $V^{\!*} \neq V$ reads
  \ba
\frac{\partial
W_{\!\psi}}{\partial
  t} + \frac{p}{m}\,\frac{\p W_{\!\psi} }{\p x}
-\frac{\p \mathrm{Re}(V)}{\p x}\,\frac{\p W_{\!\psi}}{\p p}
=2\,W_{\!\psi}\lim_{\hbar\rightarrow
0}\frac{\mathrm{Im}(V)}{\hbar}\;, \label{contLimit} \ea where
$\mathrm{Re}(V)$ and $\mathrm{Im}(V)$ represent the real and
imaginary parts of $V\!$, respectively. Unless the imaginary part
of  potential is proportional to $\hbar$, the right-hand-side in
Eq.~(\ref{contLimit}) explodes when $\hbar\rightarrow 0$. If there
is no finite limit ${\displaystyle{ \lim_{\hbar\rightarrow 0}}}
\frac{\mathrm{Im}V}{\hbar}$, we infer  from the Bohr rule that
 there is no classical system governed by a Hamiltonian $\frac{p^2}{2m}+V$.

There is an underlying assumption in formulation of Eq.~(8),  that
at every point $x$ potential $V(x)$ can  be expanded into a power
 series with an infinite radius of  convergence.
When it is not the case, relation shown in Eq.~(\ref{cont}) still
holds, but one needs to calculate $ J_p$ as a full integral
\ba
 J_p\!=\!\!  \int \!\!\! \frac{d\xi}{2\pi{\textstyle{i}}}\, e^{i\xi p/\hbar}\,
\psi^*(x\!+\!{\textstyle{\frac{\xi}{2}}})
\psi(x\!-\!{\textstyle{\frac{\xi}{2}}})
\!\bigg[\dfrac{
V\!(x\!-\frac{\xi}{2})\!-\!V\!(x)}{\xi}\!-
\!\frac{V^{\!*}\!(x\!+\frac{\xi}{2})\!-\!V^*\!(x)}{\xi}\!\bigg].
\label{e12}
\ea
Here,  both {\scriptsize{$\dfrac{
V\!(x\!-\frac{\xi}{2})\!-\!V\!(x)}{\xi} $}} and
{\scriptsize{$\dfrac{ V\!(x\!+\frac{\xi}{2})\!-\!V\!(x)}{\xi}\;$}}
 have finite values in a limit of $\xi\rightarrow 0$, which means that $J_p$
  in Eq.~(\ref{e12}) is a well defined continuous function on $(x, p)$.
Defined by Eqs. (\ref{jx}) and (\ref{e12}) Wigner function flow is
a generalization of formulas used to study phase-space dynamics in
Wigner representation \cite{Wyatt, bauke, steuer} to the case of
complex potential. Similarly, the continuity equation for the
Wigner distribution shown in Eq.~(9), along with the definition in
Eq.~(\ref{e12}), can be applied to an arbitrary complex potential,
which is not necessary Hermitian or a $\mathcal{PT}$-symmetric
one.

 For the convenience of notation, from
now on we set $\hbar=1$.  We  calculate the Wigner function
flow~$\bar{J}_{\psi}$ for the eigenstates of Eq. (1)  using a
potential $V=\frac{V_\epsilon(x)}{2}$ defined in Eq. (2). In Fig.
3, the streamline plots of $\;\bar{J}_{\!\psi}$ corresponding to
the $1$st- and $2$nd-excited states are depicted, again for
$\epsilon = 2.0$, $1.5$, and $1.4$, respectively. In the figures,
the Wigner distribution is plotted as the background.  A color of
the arrows depends on the norm
$N_{\!{}_{J}}=\sqrt{J_x^2+J_p^2}\;$, varying from white-color (for
$N_{\!{}_{J}}=0$) to dark-color (for large $N_{\!{}_{J}}$).
 As
illustrated in Fig. 3(a) and (d),   streamlines corresponding to the
harmonic oscillator  form perfect circles. The streamlines rotate
clockwise or counterclockwise, depending on the sign of the Wigner
function.   In general, the sign of $J_x$ depends on the sign of the
Wigner function and momentum: $\mathrm{sgn}(J_x)=\mathrm{sgn}(p
W_{\!\psi})$, and $\;\bar{J}_{\!\psi}$ vanishes at the points where
the Wigner function vanishes.

As demonstrated in Fig. 3,  for $\epsilon < 2$ the streamlines are
no longer perfect circles. However, the deformation in the
streamlines just reflects  changes in the Wigner function, but the
characteristics of the field is not alerted qualitatively. As long
as the $\mathcal{PT}$-symmetry is unbroken, these streamlines still
form closed loops  and their orientation depends on the sign of the
Wigner distribution, see Fig. 3(b) and (e). However, when the
$\mathcal{PT}$-symmetry is broken, as shown in Fig. 3(c) and (f), a
quantitative difference will be revealed for these Wigner function
currents by the means of Gauss-Ostrogradsky theorem.

\begin{figure}
\begin{center}
\includegraphics[width=10cm]{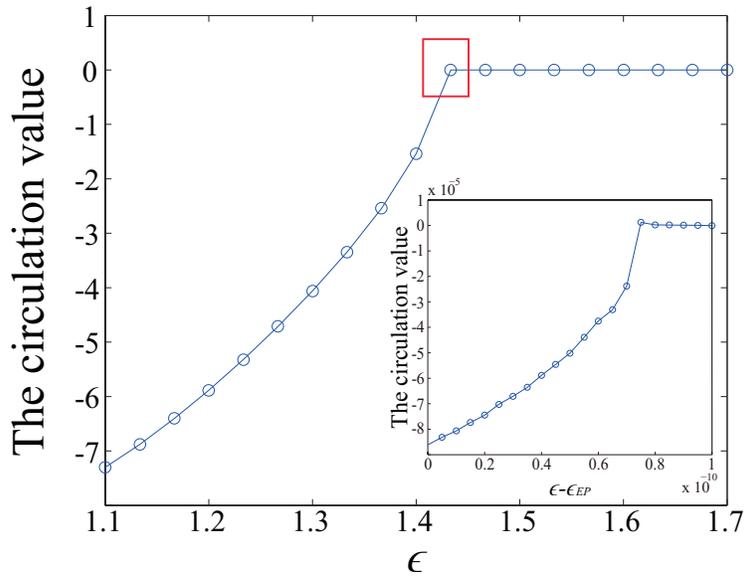}
\end{center}
\caption{The circulation value, i.e., the integral  calculated  by
Eq. (14) for the  Wigner function flow from the $1$st-excited  as
a function of the parameter $\epsilon$. Markers in circles denote
numerically values. Here, the exceptional point happens at
$\epsilon_{EP} \approx 1.42207$. One can see that integral is
equal to zero when $\epsilon \ge \epsilon_{EP}$. Inset shows the
enlarged area in the vicinity of the exception point.
}\label{fig5}
\end{figure}

\section{Wigner Function Flow at the exceptional points}
The divergence theorem, i.e., Gauss-Ostrogradsky theorem,  states
that the flux of a vector field through a closed surface is equal
to the volume integral of the divergence over the region inside
the surface \cite{Tromba,Jackson}. It allows us to calculate the
flux through volume or surface integrals  and quantitatively
distinguishes cases of unbroken and broken
$\mathcal{PT}$-symmetry. To apply the divergence theorem, we
introduce  a three dimensional (3D) field
$\bm{\mathcal{J}_{\!\!\scriptscriptstyle{3}\,}}\defeq (
W_{\!\psi}, J_x, J_p)$, and rewrite the continuity equation for
the Wigner function in Eq. (9) by this 3D field. In this notation,
Eq.~(\ref{cont}) is expressed simply as \ba
  \nabla \cdot
  \bm{\mathcal{J}_{\!\!\scriptscriptstyle{3}\,}}
 = 2 W_{\!\psi}  \mathrm{Im}V \,.\n
\ea

From the divergence theorem we infer that a flux
${\bm{\psi_{\mathcal{J}_{\!\scriptscriptstyle{3}}}}} $ of the
field $\bm{\mathcal{J}_{\!\!\scriptscriptstyle{3}}}$  through a surface $S$ enclosing  volume
${\mathrm{U}}$ can be calculated by taking a volume integral of
 $\, 2W_{\!\scriptstyle{\Psi}} \mathrm{Im}V$ over ${\mathrm{U}}$:
  \ba
{\bm{\psi_{\mathcal{J}_{\!\scriptscriptstyle{3}}}}}=\oiint_{\mathrm{S}}
\bm{\mathcal{J}_{\!\!\scriptscriptstyle{3}}}d\mathrm{S} =
\iiint_{\mathrm{U}} (\nabla \cdot \bm{\mathcal{J}_{\!\!\scriptscriptstyle{3}}}) dtdxdp =
2\iiint_{\mathrm{U}} \mathrm{Im}V\, W_{\!\scriptstyle{\Psi}}  \,
dtdxdp\,.
\label{J3}
\ea
From Eq.~(\ref{J3}), it is clear  that the flux
${\bm{\psi_{\mathcal{J}_{\!\scriptscriptstyle{3}}}}}=0\,$ when the
imaginary part of $V$ vanishes. It follows, that the flux
${\bm{\psi_{\mathcal{J}_{\!\scriptscriptstyle{3}\,}}}}$ vanishes for
all Hermitian Hamiltonians.

Whenever the Wigner function does not depend on time, i.e.,
$ \frac{\partial W\!(x,p,t)}{\partial  t}=0$,
a flux~${\bm{\psi_{\mathcal{J}_{\!\scriptscriptstyle{3}}}}} $
of the field~$\bm{\mathcal{J}_{\!\!\scriptscriptstyle{3}}}$
 through a surface $S$ for a unit time can be calculated as a
 two-dimensional (2D) integral
$\iint 2 W_{\!\scriptstyle{\Psi}} \mathrm{Im}(V)  dxdp$.
Here, a product of the Wigner distribution and the imaginary part
 of potential~$V$, can be viewed as a  charge/probability density determining
 behavior of a field~${\bm{{\mathcal{J}_{\!\scriptscriptstyle{3}}}}} $.
Note that,  as long as $V$ is antisymmetric under the
transformation $x\rightarrow -x$, the corresponding value of
$W_{\!\scriptstyle{\Psi}} \mathrm{Im}(V) $ is zero whenever the
Wigner function is symmetric under the same transformation. It
also shows that a flux
${\bm{\psi_{\mathcal{J}_{\!\scriptscriptstyle{3}}}}} $ is non-zero
only under the presence of a non-Hermitian part of  Hamiltonian
and only when a $\mathcal{PT}$-symmetry of a given solution is
broken. This fact provides a quantitative measure, which allows us
to distinguish the cases of broken  and unbroken $\mathcal{PT}$
symmetry.

Alternatively, instead of treating time parameter $t$ as a third dimension,
one can also stay in the 2D case by defining an auxiliary field
$ {\bm{\mathcal{I}_2}} = (- J_p, J_x)$.
Now, let us consider a surface $D$ that has a boundary $C=\p D $.
In 2D,  the circulation of a field ${\bm{\mathcal{I}_2}} $ along a curve $C$
can be calculated as an area integral over $D$.
Again, from the  divergence theorem we know that a circulation from the field
 ${\bm{\psi_{{\bm{\mathcal{I}_2}} }}} $ of the field~${\bm{\mathcal{I}_2}} $
 along~$C$ can be calculated as a two-dimensional  area integral of
 $ \mathrm{Im}(V) W\!(x,p,t) -\frac{\partial W\!(x,p,t)}{\partial  t}$
 over $D$, or equivalently:
\ba
\oint_{C} {\bm{\mathcal{I}_2}} \,dC &=&
\int\!\!\!\!\int_{\!D} \left(\frac{\p \mathcal{I}_x}{\p p}
 -\frac{\p \mathcal{I}_p}{\p x}\right) dxdp =
\int\!\!\!\!\int_{\!D}\! \left(\!2W_{\!\scriptstyle{\Psi}}\mathrm{Im}V
-\frac{\partial W_{\!\scriptstyle{\Psi}}}{\partial
  t}\!\right)\! dxdp\,. \label{circ}
\ea

Figure 4 shows the circulation value, the integral from Eq.~(14)
versus the parameter $\epsilon$ for the $1$st-excited state of our
modal system with $\mathcal{PT}$-symmetry in Eqs.~(1-2). Here, the
curve $C$ encircles the area $D$, which is taken large enough to
ensure that the integral value does not change for any more
expansion. One can find that when $\epsilon \ge \epsilon_{EP}
\approx 1.42207$, the circulation value is always zero, no matter
the symmetry in the Wigner function distribution sustains shown in
Fig. 2. However, across this exceptional point, $\epsilon_{EP}$,
the circulation value has a non-zero term, which reflects a broken
$\mathcal{PT}$-symmetry phase. Moreover, we check this circulation
value in the vicinity of $\epsilon_{EP}$ to $10$th decimal places,
as shown in the Insect of Fig. 4. With this numerical check, we
can confirm that the phase transition in our
$\mathcal{PT}$-symmetric system is a continuous function of the
parameter $\epsilon$, which implies a second-order phase
transition.

\section{Conclusions}
We present a phase-space study of a non-Hermitian system deriving
a continuity equation for the Wigner distribution and arbitrary
complex potential,  defining a Wigner function flow accordingly.
In particular, we reveal  how a $\mathcal{PT}$ symmetry-breaking
manifests itself in the phase-space representation. A quantitative
measure  on the circulation value for the Wigner function flow
shows that the phase transition in the vicinity of exception point
(EP) is a continuous function of the system parameter. Our study
in phase space representation indicates that a
$\mathcal{PT}$-symmetric phase transition is a second-order phase
transition.

\section*{ACKNOWLEDGMENTS}

This work is supported in part by the Ministry of Science and
Technologies, Taiwan, under the contract No. 101-2628-M-007-003-MY4,
 No.  103-2221-E-007-056, and No. 103-2218-E-007-010.

\appendix
\renewcommand\thesection{A}
\setcounter{equation}{0}
\section*{Appendix A} \label{appA}
In this Appendix, we provide the formula for the Lauricella
hypergeometric function shown in Eq. (4). With the basic
properties of Hermite polynomials, it is easy to check that \ba
-2\frac{\partial^2 }{\partial x^2} \, |n\ran
=(2n+1)|n\ran-\sqrt{n(n-1)} \,|n-2\ran -\sqrt{(n+1)(n+2)}\,
|n+2\ran\,. \ea To calculate $ \lan m| V_\epsilon(x)|n\ran$ it is
useful to note that depending on the parity of $n$ and $m$ \ba
\int_{-\infty}^0
 e^{-x^2} H_m(x) H_n(x) |x|^{\textstyle{\epsilon}}dx=\pm \int_0^{\infty}
   e^{-x^2} H_m(x) H_n(x) |x|^{\textstyle{\epsilon}} dx\n\;.
\ea
Thus, for a given $\epsilon$,  the value of an integral
\ba
 \int_{{}_{\textstyle\mathbb{R}} }
\!\! e^{-x^2}\! H_m(x)  H_n(x)V_\epsilon(x)\,dx &=&\int_{-\infty}^0
 e^{-x^2} H_m(x) H_n(x) |x|^{\textstyle{\epsilon}}\big[\cos(\frac{\pi}{2} {\textstyle{\epsilon}})
                         -i \sin(\frac{\pi}{2} {\textstyle{\epsilon}})\big]dx+
                         \n\\
                         &+&\int_0^{\infty}
   e^{-x^2} H_m(x) H_n(x) |x|^{\textstyle{\epsilon}}\big[\cos(\frac{\pi}{2} {\textstyle{\epsilon}})
                         +i \sin(\frac{\pi}{2} {\textstyle{\epsilon}})\big]dx \n
\ea
is determined by an integral (\ref{Erd}) defined below.
\noindent
It was shown by Erd\`elyi~\cite{Erdelyi}
that{\small
\begin{eqnarray}
2\int_0^\infty&&\!\!\!\!\!\!e^{-x^2} H_{\mu}(x) H_{\nu}(x)
x^{\textstyle{\epsilon}} dx  =\n
  h_{\mu} h_{\nu}
\Gamma\big({\textstyle{1+\frac{\textstyle{\epsilon}}{2}
-\frac{(-1)^{\mu}+(-1)^{\nu}}{4} }}\big)\times\\ &&\times\,
F_A\big({\textstyle{ 1+\frac{\textstyle\epsilon}{2}
-\frac{(-1)^{\mu}+(-1)^{\nu}}{4};
\frac{ \{\mu\} - \mu}{2}  ;
\frac{ \{\nu\} - \nu}{2}  ;
1-\frac{(-1)^{\mu}}{2};  1-\frac{(-1)^{\nu}}{2}; 1,1 }}\big)\,,
\label{Erd}
\end{eqnarray}
}where $\Gamma(.)$ is an Euler gamma function, $F_A(.)$ denotes a
Lauricella hypergeometric function and parameters $h_{\mu}$,
$h_{\nu}$  are defined as:
\ba
h_{\mu} =\begin{cases}
   (-1)^\frac{\mu}{2} \,\mu! \left[\left(\frac{\mu}{2}\right)!\right]^{-1};&
   \text{for even} \;\mu  \\
   (-1)^\frac{\mu+1}{2}\, 2\mu! \left[\left(\frac{\mu-1}{2}\right)!\right]^{-1};&
   \text{for odd} \;\mu
  \end{cases}\nonumber
\ea
For different parities of $\mu$ and $\nu$ one obtains:
{\small{\begin{eqnarray} &&\!\!\!\!\!\!\!\!\int_0^\infty
\!\!\!e^{-x^2} H_{2r}(x\!) H_{2\!s}(x\!)\, x^{\textstyle{\epsilon}}dx =
\frac{(-1)^{r+s} (\!2r\!)!(\!2s\!)!  }{2 r! s!}  \,
{\displaystyle{\Gamma}}\!\left(\textstyle{\frac{{\textstyle{\epsilon}}+1}{2}}
\right) F_A\!\!\left( {\textstyle{\frac{{\textstyle{\epsilon}}+1}{2}}}\,;
-r  , -s \,;{\textstyle \frac{1}{2}},
{\textstyle \frac{1}{2}}\,; 1,1
\right),
\label{Er_1}
\\
&&\!\!\!\!\!\!\!\!\int_0^\infty \!\!\! e^{-x^2} H_{2r}(x\!)
H_{2\!s+1}\!(x\!)
\,x^{\textstyle{\epsilon}}dx = \frac{(-1)^{r+s} (\!2r\!)!(\!2s+1\!)!  }{ r!
s!}  \,
\Gamma\!\left(\textstyle{\frac{{\textstyle{\epsilon}}+2}{2}}
\right) F_A\!\!\left( {\textstyle{\frac{{\textstyle{\epsilon}}+2}{2}}}\,;
-r  , -s \,;{\textstyle \frac{1}{2}},
{\textstyle \frac{3}{2}}\,; 1,1
\right),
\label{Er_2}
\\
&&\!\!\!\!\!\!\!\!\int_0^\infty\!\!\!  e^{-x^2} H_{2r+1}\!(x\!)
H_{2\!s+1}\!(x\!)\, x^{\textstyle\epsilon}dx= \frac{2(-1)^{r+s}
(\!2r+1\!)!(\!2s+1\!)!  }{ r! s!}  \,
\Gamma\!\left(\textstyle{\frac{{\textstyle{\epsilon}}+3}{2}}
\right) F_A\!\!\left( {\textstyle{\frac{{\textstyle{\epsilon}}+3}{2}}}\,;
 -r  , -s \,;{\textstyle \frac{3}{2}},
{\textstyle \frac{3}{2}}\,; 1,1
\right),\n
\label{Er_3}
\end{eqnarray}}}Relation (\ref{Erd}) is a special case of more general formula
for an integral $\int_0^{\infty}
   e^{-\lambda x^2} H_{\mu_1}\!(\beta_1 x) H_{\mu_2}\!(\beta_2 x)\cdots H_{\mu_n}\!(\beta_n x)
    x^{\nu} dx$, which we don't rewrite here because of its length.
 Lauricella hypergeometric function is defined as
\begin{eqnarray}
F_A^{(n)}(a;b_1,...,b_2;c_1,...,c_n;x_1,...,x_n)=\!\!\!\!\!\!\sum_{i_1,...,i_n=0}^\infty
\!\!\!\frac{ \bm{(}a\bm{)}_{i_1+...+i_n}
{\bm{(}}b_1\bm{)}_{i_1}\cdots  \bm{(}b_n\bm{)}_{i_n} }{
 \bm{(}c_1\bm{)}_{i_1}\cdots \bm{(}c_n\bm{)}_{i_n} i_n!
\cdots i_n!}\, x_1^{i_1}\cdots x_n^{i_n}
\end{eqnarray}
with $\,{\bm{(}} .\bm{)}_{k}\,$ being a Pochhammer symbol, i.e.,
$\,{\bm{(}}a\bm{)}_{k}=a(a+1)\cdots (a+k-1)$. In our case, $n=2$ and
$F_A(.)$ are of the form
\ba
F_A\left( {\textstyle{\frac{{\textstyle{\epsilon}}+1}{2}}}\,; -r  ,
-s \,;{\textstyle \frac{1}{2}},  {\textstyle \frac{1}{2}}\,; 1,1
\right)\n=
\sum_{i_1,i_2=0}^\infty
\frac{ \bm{(}{\textstyle{\frac{{\textstyle{\epsilon}}+1}{2}}}\bm{)}_{i_1+i_2}
{\bm{(}}-r\bm{)}_{i_1} \bm{(}-s\bm{)}_{i_2} }{
\bm{(}\frac{1}{2}\bm{)}_{i_1} \bm{(}\frac{1}{2}\bm{)}_{i_2} \,i_1! \,
i_2!}\, .
\ea
Because $r$ and $s$ are natural numbers there will be always a
finite number of terms in the sum as ${\bm{(}}-r\bm{)}_{k} = 0\quad
$ for $\,k\geq r+1\n\,,$ so
\ba
&&\sum_{i_1,i_2=0}^\infty
\frac{ \bm{(}{\textstyle{\frac{{\textstyle{\epsilon}}+1}{2}}}\bm{)}_{i_1+i_2}
{\bm{(}}-r\bm{)}_{i_1} \bm{(}-s\bm{)}_{i_2} }{
\bm{(}\frac{1}{2}\bm{)}_{i_1} \bm{(}\frac{1}{2}\bm{)}_{i_2} \,i_1! \,
i_2!} = \sum_{i_1,i_2=0}^{r,s}
\frac{ \bm{(}{\textstyle{\frac{{\textstyle{\epsilon}}+1}{2}}}\bm{)}_{i_1+i_2}
{\bm{(}}-r\bm{)}_{i_1} \bm{(}-s\bm{)}_{i_2} }{
\bm{(}\frac{1}{2}\bm{)}_{i_1} \bm{(}\frac{1}{2}\bm{)}_{i_2} \,i_1! \,
i_2!}\, =\n\\
&&=1+\sum_{i_1=1}^{r}
\frac{ \bm{(}{\textstyle{\frac{{\textstyle{\epsilon}}+1}{2}}}\bm{)}_{i_1}
{\bm{(}}-r\bm{)}_{i_1} }{
\bm{(}\frac{1}{2}\bm{)}_{i_1}  \,i_1! }\,
+
\sum_{i_2=1}^{s}
\frac{ \bm{(}{\textstyle{\frac{{\textstyle{\epsilon}}+1}{2}}}\bm{)}_{i_2}
{\bm{(}}-s\bm{)}_{i_1} }{
\bm{(}\frac{1}{2}\bm{)}_{i_2}  \,i_2! }\,+\sum_{i_1,i_2=1}^{r,s}
\frac{ \bm{(}{\textstyle{\frac{{\textstyle{\epsilon}}+1}{2}}}\bm{)}_{i_1+i_2}
{\bm{(}}-r\bm{)}_{i_1} \bm{(}-s\bm{)}_{i_2} }{
\bm{(}\frac{1}{2}\bm{)}_{i_1} \bm{(}\frac{1}{2}\bm{)}_{i_2} \,i_1! \,
i_2!}.\,\n
\ea
We can use following formula:
\ba
{\bm{(}}-r\bm{)}_{k}=(-r)(-r+1)\cdots (-r+k-1)  =(-1)^k\, r\,(r-1)
\cdots (r-(k-1))= \frac{(-1)^k\,r!}{(r-k)!},  \n
\ea
and rewrite
\ba
\sum_{i_1=1}^{r}
\frac{ \bm{(}{\textstyle{\frac{{\textstyle{\epsilon}}+1}{2}}}\bm{)}_{i_1}
{\bm{(}}-r\bm{)}_{i_1} }{
\bm{(}\frac{1}{2}\bm{)}_{i_1}  \,i_1! }\,= r!\sum_{i_1=1}^{r}
\frac{ (-1)^{i_1}  \bm{(}{\textstyle{\frac{{\textstyle{\epsilon}}+1}{2}}}\bm{)}_{i_1}
}{
\bm{(}\frac{1}{2}\bm{)}_{i_1}  \,i_1! (r-i_1)!}\,=
r!\sum_{i_1=1}^{r}
\frac{ (-1)^{i_1} \prod_{k=1}^{i_1}(\epsilon+2k-1)
}{ (2i_1-1)!!  \,i_1! \, (r-i_1)!},\,\label{e27}
\ea
where we use also the facts that:
\ba
{\bm{\Big(}}\frac{1}{2}\bm{\Big)}_{n}&=& \frac{(2n-1)!!}{2^n},\n
\ea
and
\ba
{\bm{\Big(}}\frac{\epsilon+1}{2}\bm{\Big)}_{n}&=&
\frac{(\epsilon+1)(\epsilon+3)\cdots (\epsilon+2n-1)}{2^{n}}   =
\frac{1}{2^{n}} \, {\displaystyle\prod_{k=1}^n(\epsilon+2k-1)}.\n
\ea
Similarly, we have
\ba
\sum_{i_2=1}^{s}
\frac{ \bm{(}{\textstyle{\frac{{\textstyle{\epsilon}}+1}{2}}}\bm{)}_{i_2}
{\bm{(}}-s\bm{)}_{i_2} }{
\bm{(}\frac{1}{2}\bm{)}_{i_2}  \,i_2! }\,=
s!\,\sum_{i_1=1}^{s}
\frac{ (-1)^{i_2} \prod_{k=1}^{i_2}(\epsilon+2k-1)
}{ (2i_2-1)!!  \,i_2! \, (r-i_2)!},\,\label{e28}
\ea
and
\ba
\sum_{i_1,i_2=1}^{r,s}
\frac{ \bm{(}{\textstyle{\frac{{\textstyle{\epsilon}}+1}{2}}}\bm{)}_{i_1+i_2}
{\bm{(}}-r\bm{)}_{i_1} {\bm{(}}-s\bm{)}_{i_2} }{
\bm{(}\frac{1}{2}\bm{)}_{i_1} \bm{(}\frac{1}{2}\bm{)}_{i_2}  \,i_1\,i_2! }\,=
s!\,\sum_{i_1=1}^{s}
\frac{ (-1)^{i_2} \prod_{k=1}^{i_2}(\epsilon+2k-1)
}{ (2i_2-1)!!  \,i_2! \, (r-i_2)!}.\,\label{e29}
\ea
Combining Eqs. (\ref{e27}), (\ref{e28}) and (\ref{e29}) we can
rewrite $F_A\left(
{\textstyle{\frac{{\textstyle{\epsilon}}+1}{2}}}\,; -r  , -s
\,;{\textstyle \frac{1}{2}},  {\textstyle \frac{1}{2}}\,; 1,1
\right) $ in a convenient form of finite sums.

Using facts that
\ba
&&{\bm{\Big(}}\frac{3}{2}\bm{\Big)}_{n}=\frac{(2n-1)!!}{2^{n+1}}
\,,\qquad \mathrm{and}
\n\\
&&{\bm{\Big(}}\frac{\epsilon+2}{2}\bm{\Big)}_{n}=
\frac{(\epsilon+2)(\epsilon+4)\cdots (\epsilon+2n)}{2^{n}}   =
\frac{1}{2^{n}} \, {\displaystyle\prod_{k=1}^n(\epsilon+2k)}\,,\n\\
&&{\bm{\Big(}}\frac{\epsilon+3}{2}\bm{\Big)}_{n}=
\frac{(\epsilon+3)(\epsilon+5)\cdots (\epsilon+2n+1)}{2^{n}}   =
\frac{1}{2^{n}} \, {\displaystyle\prod_{k=1}^n(\epsilon+2k+1)}\,,\n
\ea
analogues formulas  for two other functions $F_A \Big(
{\textstyle{\frac{{\textstyle{\epsilon}}+2}{2}}}\,; -r  , -s
\,;{\textstyle \frac{1}{2}},  {\textstyle \frac{3}{2}}\,; 1,1
\Big)$   and $ F_A\Big( {\textstyle{\frac{{\textstyle{\epsilon}}+3}{2}}}\,; -r  ,
-s \,;{\textstyle \frac{3}{2}},  {\textstyle \frac{3}{2}}\,; 1,1
\Big)\n$ are obtained. Together they simplify
 calculation of elements $a_{nm}(\epsilon)$,Eq. (4), considerably. In
principle, finding these elements can be done by hand without any
aid of numerics.


\begin{thebibliography}{99}
\bibitem{comment1}
To define an operator one has to state explicitly its domain
(usually a vector space), to talk about hermiticity also a
definition of an inner product on this space is required.
The spectral theorem makes a classification of finite-dimensional
cases fairly simple: if an operator $\hat{M}$ is diagonalizable
and has purely real eigenvalues, then there exists a scalar
product  in respect to which $\hat{M}$ is Hermitian. If an
operator $\hat{M}$ is diagonalizable but has some non-real
eigenvalues, it can always be decomposed into a sum of two,
mutually commuting, hermitian operators. In the
infinite-dimensional case, the linear independence of vectors is
no longer sufficient to guarantee that a mapping between basis
corresponding to different scalar products is continuous.  A much
stronger condition that there are no limit points in the set of
eigenvectors has to be fulfilled.
\end{thebibliography}
\end{document}